\documentclass[11pt]{article}
\usepackage{times}
\usepackage{geometry}
\usepackage[utf8]{inputenc}
\usepackage{enumitem,amssymb}
\usepackage{graphicx}
\usepackage{fancyhdr}
\usepackage{aas_macros}
\usepackage{mdframed} 
\usepackage{pdfpages}
\usepackage{url}
\usepackage[authoryear]{natbib}
\usepackage{wrapfig}
\usepackage{ragged2e}

\newcommand\blfootnote[1]{%
	\begingroup
	\renewcommand\thefootnote{}\footnote{#1}%
	\addtocounter{footnote}{-1}%
	\endgroup
}

\setcitestyle{authoryear,open={(},close={)}}
\mdfdefinestyle{theoremstyle}{
innertopmargin=\topskip,}
\mdtheorem[style=theoremstyle]{lrptextbox}{}

\graphicspath{{figures/}}

\begin{document}




%

\section*{\huge LRP2020: Astrostatistics in Canada}

\section*{Authors}

Gwendolyn Eadie\blfootnote{PI, corresponding author: gwen.eadie@utoronto.ca}\footnotemark[1]$^,$\footnotemark[2]$^,$\footnotemark[3]$^,$\footnotemark[4], Arash Bahramian\footnotemark[5], Pauline Barmby\footnotemark[6], Radu Craiu\footnotemark[2], Derek Bingham\footnotemark[7]$^,$\footnotemark[8], Ren\'ee Hložek\footnotemark[1]$^,$\footnotemark[9], JJ Kavelaars\footnotemark[10], David Stenning\footnotemark[11], Samantha Benincasa\footnotemark[12], Guillaume Thomas\footnotemark[10], Karun Thanjavur\footnotemark[13], Jo Bovy\footnotemark[1]$^,$\footnotemark[9], Jan Cami\footnotemark[6], Ray Carlberg\footnotemark[1], Sam Lawler\footnotemark[10]; Adrian Liu\footnotemark[14]$^,$\footnotemark[15], Henry Ngo\footnotemark[10], Mubdi Rahman\footnotemark[9], Michael Rupen\footnotemark[10]

\footnotetext[1]{Dept. of Astronomy \& Astrophysics, University of Toronto, Toronto. ON}
\footnotetext[2]{Dept. of Statistical Sciences, University of Toronto, Toronto, ON}
\footnotetext[3]{American Astronomical Society Working Group on Astrostatistics \& Astroinformatics, USA}
\footnotetext[4]{American Statistical Association Astrostatistics Interest Group, USA}	
\footnotetext[5]{International Center for Radio Astronomy Research, Curtin University, Australia}
\footnotetext[6]{Western University, London, ON}
\footnotetext[7]{Dept. of Statistics \& Actuarial Science, Simon Fraser university, Burnaby, BC}
\footnotetext[8]{Canadian Statistical Science Institute Headquarters, Burnaby, BC}
\footnotetext[9]{Dunlap Institute for Astronomy \& Astrophysics, University of Toronto, Toronto, ON}
\footnotetext[10]{NRC Herzberg Astronomy \& Astrophysics Research Center, Saanich, BC}
\footnotetext[11]{Dept. of Mathematics, Imperial College London, London, UK}
\footnotetext[12]{Dept. of Physics, University of California Davis, Davis, CA}
\footnotetext[13]{Dept. of Physics \& Astronomy, University of Victoria, Victoria, BC}
\footnotetext[14]{Dept. of Physics, McGill University, Montreal, QC}
\footnotetext[15]{Institut Spatial de McGill/McGill Space Institute, McGill University, Montreal, QC}


\section*{Executive Summary}
This state-of-the-profession white paper focuses on the interdisciplinary fields of astrostatistics and astroinformatics, in which modern statistical and computational methods are applied to and developed for astronomical data. Astrostatistics and astroinformatics have grown dramatically in the past ten years, with international organizations, societies, conferences, workshops, and summer schools becoming the norm. Canada’s formal role in astrostatistics and astroinformatics has been relatively limited, but there is a great opportunity --- and necessity --- for growth in this area.

In the 2020s, extremely large astronomy datasets will be available from both Canadian- and internationally-funded projects and missions. Millions of dollars and thousands of human hours have been invested in order to obtain these data, and we need to make the most of these data when performing scientific inference. Novel statistical and computational methods from astrostatistics and astroinformatics will be the driving force in the next decade of scientific discovery, and interdisciplinary collaboration is key.

In order to establish Canada as an international leader in astrostatistics and astroinformatics, we must first understand our current state in these areas. Thus, we conducted a survey of astronomers in Canada to gain information on the training mechanisms through which we learn statistical methods and to identify areas for improvement. We explore Canadian-based astronomers'  exposure to and training in statistical methodology, their self-described statistical acumen, and their suggestions for improving training in statistics in astronomy programs and for increasing interdisciplinary collaboration with statisticians. In general, the results of our survey indicate that while astronomers see statistical methods as critically important for their research, they lack focused training in this area and wish they had received more formal training during all stages of education and professional development. These findings inform our recommendations for the Long Range Plan 2020 on how to increase interdisciplinary connections between astronomy and statistics at the institutional, national, and international levels over the next ten years. We recommend specific, actionable ways to increase these connections, and discuss how interdisciplinary work can benefit not only research but also astronomy's role in training Highly Qualified Personnel (HQP) in Canada.

\newpage

\newpage
\section{Introduction}

The rate of data collection in astronomy has increased substantially in the past ten years with the advent of many ground-based telescopes and space-based projects like the the Atacama Large Millimeter Array (ALMA), the Gaia satellite, the Kepler Spacecraft, the Transiting Exoplanet Survey Satellite (TESS), and the Canadian Hydrogen Intensity Mapping Experiment (CHIME). The increasing rate of data collection will continue as big data projects such as the Large Synoptic Survey Telescope (LSST), the Square Kilometer Array (SKA), and the Euclid space observatory come online in the 2020s. Most of these projects will produce large, publicly available data sets that can be accessed by astronomers around the world. Millions of dollars and thousands of human hours are invested in building these kinds of projects so that we can better understand the Universe. Therefore, it is imperative that astronomers and astrophysicists strive to obtain the most information possible from these data and to perform scientific inferences through the appropriate use of statistical and data-analytical tools. 


Not surprisingly, the use of statistical methods for astronomy and astrophysics research is ubiquitous and continues to accelerate. To date, a search in Google Scholar for ``astronomy and statistics" yields 2 million hits! However, focused training in statistics is generally not emphasized in astronomy curricula. This issue has been acknowledged for over ten years, and positive changes have been taking place with the growth of fields such as Astrostatistics and Astroinformatics. 

\textbf{Astrostatistics and Astroinformatics are interdisciplinary fields that perform research at the interface of astronomy and statistics, computer science, applied math, and data analytics.} In the past ten years, these interdisciplinary fields have shown rapid growth around the world, in part due to the recognized importance of statistical analysis at all stages of astronomical research, from exploration to inference and discovery.

\subsection{The Community of Astrostatistics \& Astroinformatics}
In 2009, two white papers on astroinformatics and astrostatistics \citep{borne2009,loredo2009} were put forward for the Astro2010 Decadal Review in the United States. These papers emphasized the potential of the interdisciplinary fields to make significant contributions to astronomy, and led to the formation of the American Astronomical Society (AAS) Working Group on Astrostatistics and Astroinformatics (WGAA) in June 2012. In that same year, Joseph Hilbe (1944-2017) formed the International Astrostatistics Association (IAA), which now has over 600 members. Other active groups, including the American Statistical Association (ASA) Astrostatistics Interest Group (AIG) and the International AstroInformatics Association (IAIA), formed soon thereafter (see Table~\ref{tab:societies}). 
These groups, both within the US and internationally, have been fostering collaborations between astronomy and disciplines such as statistics, computer science, and applied math, and are advocating for educational changes to include training of statistical analyses and computational techniques in astronomy curricula.


\begin{table}[h]
    \centering
      \caption{Astrostatistics and Astroinformatics Centres, Working Groups, Associations, etc.}
    \label{tab:societies}
    \begin{tabular}{p{5cm}|p{1.35cm}|p{9cm}}
\textbf{Association/Group} & \textbf{Formed} & \textbf{Activities and/or Objective} \\
\hline
\hline
    Centre for Astrostatistics & 2003 & organise an annual Summer School in Statistics for Astronomers, astrostatistics research, hosts the Astrostatistics \& Astroinformatics Portal \small{\url{https://asaip.psu.edu/}} \\
    \hline
    LSST's Informatics \& Statistics Science Collaboration & 2009 & develop tools for large astronomical surveys \\
    \hline
    AAS WGAA  & 2012 & organise sessions and panels at AAS conferences; advocate for curricula change \\
    \hline
    IAA & 2012 & foster collaborations between statisticians and astronomers \\
    \hline
    Astrostatistics Facebook Group & 2013 & over 4300 members, active discussion \\
    \hline
    ASA AIG & 2014 & increase statisticians' involvement in astronomy research; propose and organise sessions at the annual Joint Statistical Meetings; annual student paper competition \\
    \hline
    IEEE Astrominer Task Force & 2014 & contribution to machine learning, data-mining, and computational methods \\
    \hline
    IAU Commission B3 & 2015 & Commission on Astroinformatics and Astrostatistics (previously Working Group since 2012) \\ 
    \hline
    IAIA & 2019 & host the annual astroinformatics conference
\end{tabular}
\end{table}

Through these groups and other initiatives, research activity in the fields of astrostatistics and astroinformatics has also grown dramatically, including \textbf{conferences and conferences series} (e.g. \small{\url{astroinformatics.org}}, \small{\url{www.cosmostat.org/}}, \normalsize \emph{Statistical Challenges in Modern Astronomy I-VI}, \normalsize \citealt{ford2007statistical,babu2012statistical}), \textbf{workshops and summer schools} (e.g., \emph{Penn State Summer School in Statistics for Astronomers I-XV, Stats4Astro: Bayesian Methodology 2018, ESAC Data Analysis \& Statistics Workshop 2019}),
\textbf{peer-reviewed articles}\footnote{ ``The number of articles with keyword ‘Methods: Statistical’ increased by a
factor of 2.5 in the past decade; those with ‘machine learning’ increased by 4 times over five
years; and those with ‘deep learning’ have more than tripled every year since 2015.'' -- \cite{Siemiginowska2019}. Also see the LRP2020 white paper \emph{Machine Learning Advantages in Canadian Astrophysics} by Venn et al (2019).}, \textbf{interdisciplinary research initiatives} (e.g. the COIN and SAMSI programs), \textbf{new journal series} (e.g. Cambridge University Press Elements Series on Astrostatistics), \textbf{books} \citep[e.g.,][]{hilbe2012astrostatistical,feigelson2012modern, ivezic2014statistics, hilbe2017bayesian}, \textbf{data challenges} \citep[e.g., the PLAsTiCC Astronomical Classification Challenge,][]{Hlozek2019PASP} 
, and most recently a \textbf{student paper competition} (Joint Statistical Meetings 2019, JSM 2020). Two white papers were also submitted to the Astro2020 Decadal Review this past year: the science white paper ``The Next Decade of Astroinformatics and Astrostatistics'' \citep{Siemiginowska2019} and the state-of-the-profession white paper ``Realizing the potential of astrostatistics and astroinformatics''. \citep{2019arXivEadie}


The growth of these fields is due in part to an obvious and demonstrated need for highly-qualified personnel (HQP) with strong statistical analysis and computational skills to make the most efficient use of data. Astrostatistics and astroinformatics are also receiving recognition as valuable assets to astronomy, as demonstrated by the creation and increased availability of interdisciplinary positions for postdocs and faculty in these fields (the number of jobs for Ph.D. astronomers that emphasize data analysis methodology doubled between 2013 and 2019 \footnote{\url{https://asaip.psu.edu/resources/jobs}}). 


\subsection{Connection and Relevance to Canada}
The Canadian astronomical community has a strong international presence, and has world-renowned research universities and national centers like NRC Herzberg, the Canadian Astronomy Data Centre, the Canadian Institute for Theoretical Astrophysics, and the Dunlap Institute. \textbf{Yet, almost all of the astrostatistics and astroinformatics associations and many of the activities (e.g. workshops, conferences, and summer schools) have been formed and are based outside of Canada.}

This is not to say that Canadians are not involved in astrostatistics. One Canadian-based astronomer serves on the executive committee of the ASA AIG and is co-chair of the ASA WGAA, others are members of the LSST Informatics \& Statistics Science collaboration,  and in the past year two Canadian institutions hired interdisciplinary faculty who specialize in astrostatistics. Some astronomy graduate programs in Canada also offer graduate courses in astrostatistics. \textbf{Moreover, there is a sense of demand from undergraduate and graduate students in Canada for training in modern statistical and computational techniques --- not only to help their research but also to improve their job prospects in industry.}

Some natural questions then arise about Canadian-based astronomers' views and experiences in astrostatistics and astroinformatics. For example, \emph{How important is statistical analysis in our research?}, \emph{How many of us have collaborated with a statistician, computer scientist, or applied mathematician?}, \emph{How, when, and where are we trained in statistics?}, and \emph{Do we feel this training is adequate?}. Furthermore, \emph{What can we do to improve our involvement in astrostatistics and astroinformatics?}, \emph{What can we do to engage more statisticians and computer scientists in our work?}, and \emph{What types of challenges might need to be overcome to create more interdisciplinary collaborations?}. These types of questions motivated our white paper, and led us to perform a survey on astrostatistics in Canadian-based astronomy. The survey responses then inform our recommendations to the LRP2020. For our survey and in the rest of this paper, we focus on astrostatistics and interdisciplinary collaboration with statisticians, but many of our recommendations are transferable to astroinformatics as well. 

The following outlines the rest of this paper. We start by describing the survey and presenting the results in Section~\ref{sec:survey}. In Section~\ref{sec:discussion}, we provide a discussion and include direct quotes from survey respondents to supplement the conversation. In Section~\ref{sec:recs}, we propose a number of actionable recommendations for improving the educational training in statistics for astronomers and increasing interdisciplinary collaboration with statisticians at the institutional, national, and international levels.

\section{Survey on statistical training of astronomers and interdisciplinary collaboration in Canada}\label{sec:survey}

We performed a Google survey targeting the current and former members of the Canadian Astronomical Society (CASCA), to (1) gauge their background training in statistics, (2) find out about their learning and research resources for statistics, (3) inquire about the importance of statistics in their research, and (4) ask for comments and feedback on the perceived and/or experienced barriers to interdisciplinary collaboration.  The survey was sent to the CASCA email list in July 2019, and people were encouraged to forward the survey to past students and colleagues who were trained in astronomy in Canada. A copy of the survey questions can be found at \small{\url{https://bersavosh.github.io/files/lrp_astrostat_survey.pdf}.}\normalsize

Overall, the survey contained 16 questions: two demographic questions about career stage and general field of work, ten multiple choice questions (often with the option to check all that applied and/or write in an ``other'' category), and four open response questions.  The data were collected anonymously and respondents were informed that all or part of their responses could be reproduced verbatim in this white paper. 
 
The survey and its results should be regarded as preliminary and exploratory analysis; a future, more complete survey could be conducted in consultation with an expert in survey design and analysis. We also note that many of the faculty in Canada (some of whom responded to the survey) did not obtain their degrees in Canada, so some of their responses may reflect international experiences. 

\subsection{Survey Results}

There were 123 responses to our survey; 50 faculty (including 5 retired, 4 teaching stream, and 1 non-tenured), 34 graduate students, 15 researchers in academia/government, 13 postdoctoral fellows, 6 undergraduates, and 5 non-academics. The number of CASCA members in good standing as of June 2019 was 490, making the response rate roughly 25\% (the number of non-CASCA respondents was very low). Roughly two thirds of the respondents classified their research within the observational realm, a quarter classified their research as theoretical, and the remainder indicated they were either a mixture of the two, worked in instrumentation, or were a data scientist. These demographics summarize the responses to Questions 1 and 2.

\begin{wrapfigure}{r}{0.53\textwidth}
	\centering
	\includegraphics[scale=0.61]{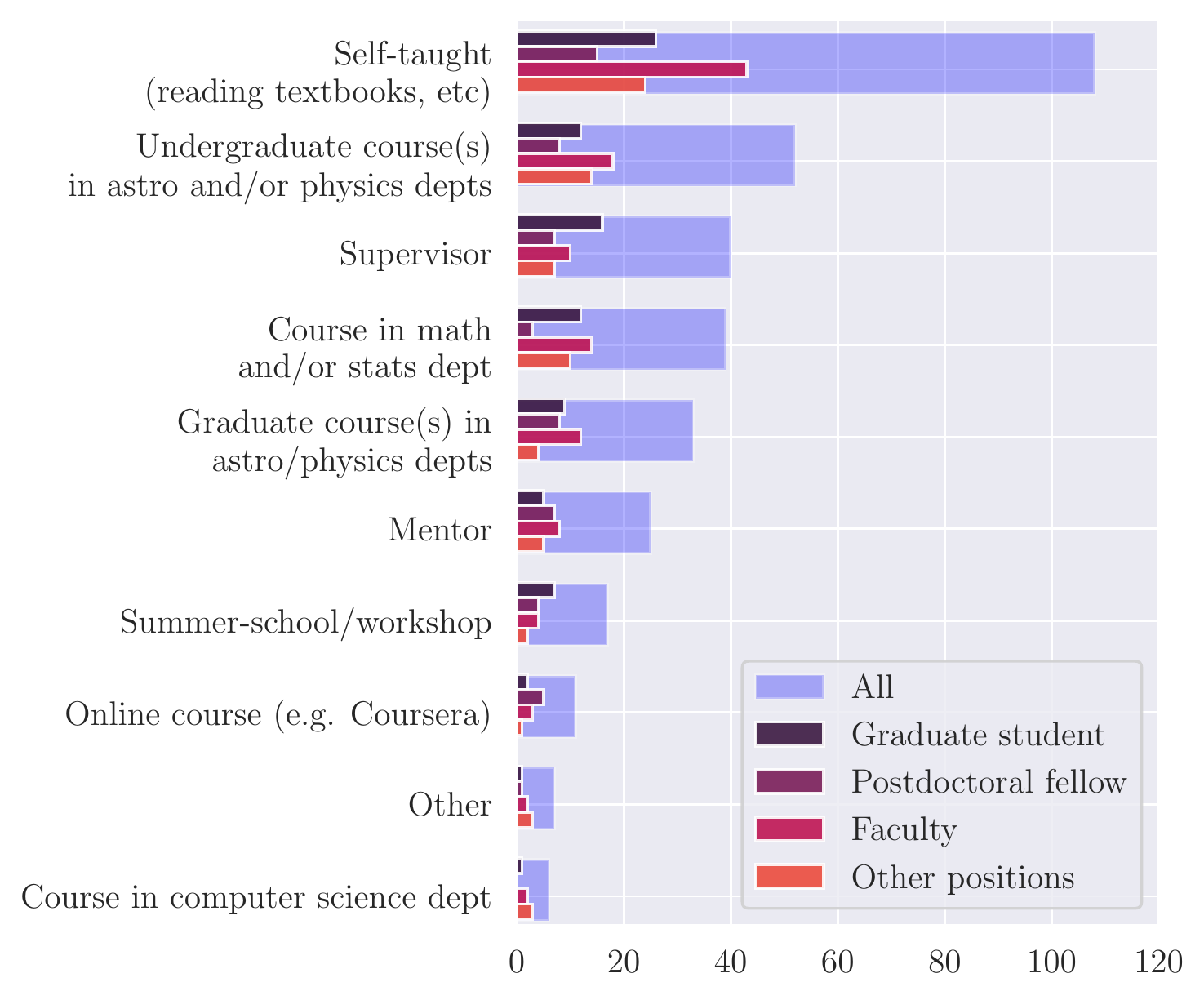}
	\caption{Results of Question 4: \emph{``When/from whom did you receive training in statistics?''} Respondents could check all that applied. Overwhelmingly, respondents indicated that they were self-taught in statistics.
	}
	\label{fig:q4}
	\vspace{-1ex}
\end{wrapfigure}

Question 3 asked respondents to rate how important statistical analysis and applied statistics was in their work, with 1 meaning ``not important'' and 5 meaning ``critically important''. \textbf{Respondents overwhelming indicated that statistics was towards the ``critically important" level, with approximately 78\% of responses being either 4 or 5 (5 was chosen by 40.7\% of all respondents).}  Question 4 then asked how people received training in statistics (Figure~\ref{fig:q4}). While the majority of respondents believe statistical analysis plays an essential role in their work, almost 85\% of respondents indicated that they were self-taught in statistics. \textbf{Most of the respondents have learned statistics by themselves or received training in an undergraduate physics or astronomy course}. 

Questions 5 and 6 asked if any of the statistics course(s) taken 
were mandatory, recommended, neither, or not applicable. Out of the respondents who took a statistics course \emph{within} their department, about 46\% said this training was mandatory. Of the respondents who took a statistics course \emph{outside} their astro/physics departments, about 28\% said the courses were mandatory.


Question 7 asked: \emph{If you currently work in industry, did you take additional training in statistics/data science after you left academia?}. Unfortunately, most of the responses were ``NA'' (116/123), with the remainder almost evenly split between ``yes'' (3) and ``no'' (4). We therefore cannot draw any conclusions here.

Question 8 asked how much training in statistics was received within Canada. Roughly 38\% of respondents received all of their training in Canada whereas roughly 43\% received none of their training in Canada, and the rest fell somewhere in between. However, this is perhaps not very informative; faculty represent about 40\% of the survey participants, and many of these people may have received their degrees outside of Canada.

\textbf{When respondents were asked if they wished they had received more training in statistics at the undergraduate, graduate, and postdoc levels} (i.e. Question 9), \textbf{the number of ``yes'' responses were 105 (85\%), 101 (82\%), and 45 (36\%) respectively}. For the postdoc-level responses, 53/123 respondents selected ``not applicable'' (i.e. they are not at this career stage or didn't do a postdoc). For the rest of the postdoc-level responses, approximately 65\%  wished they had more training in statistics.

\begin{figure}[h]
\vspace{-1.5ex}
    \centering
    \includegraphics[scale=0.48, trim={0cm 1 0 1}, clip]{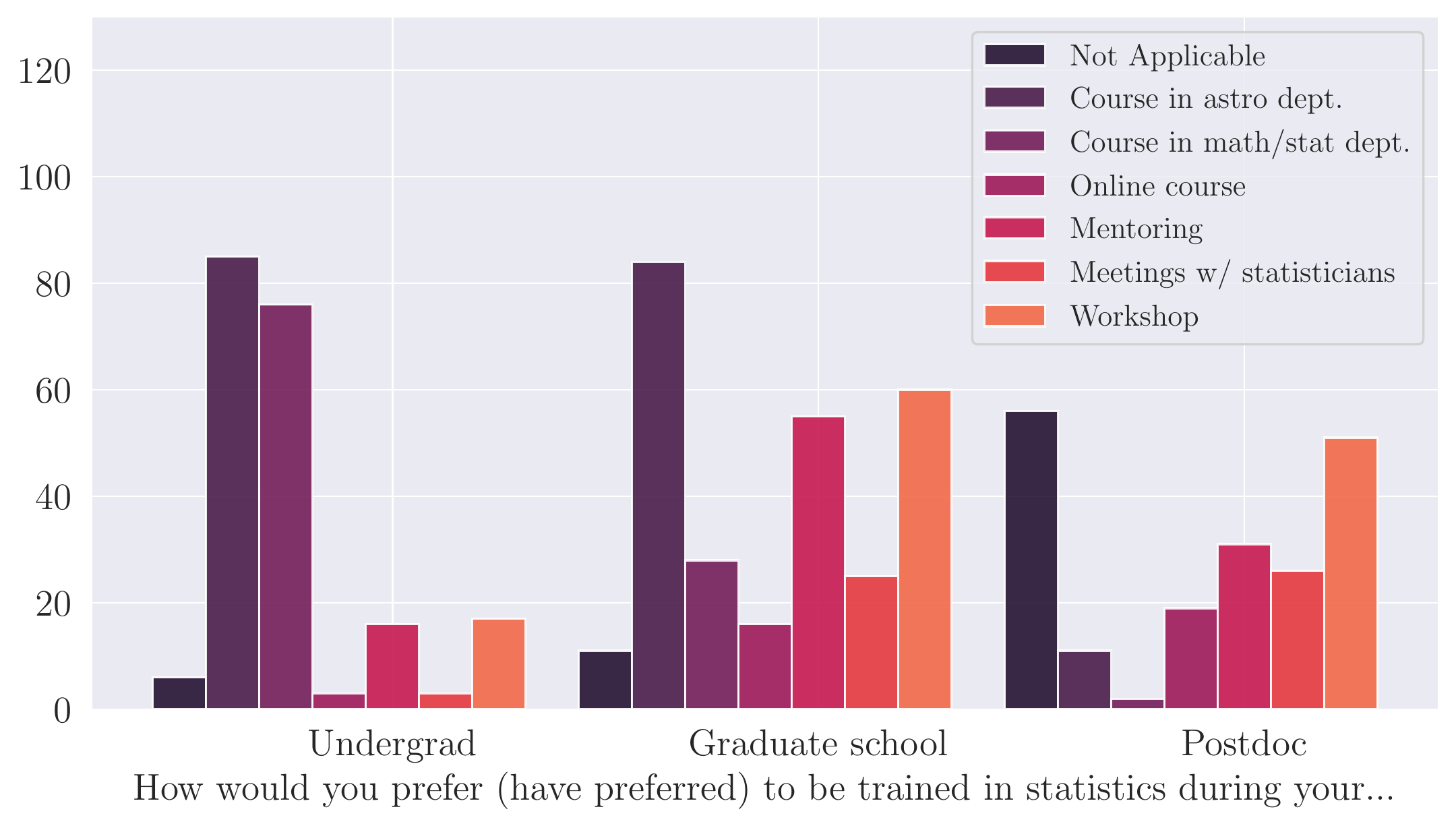}
    \caption{Results of Question 10: \emph{``How would you prefer (have preferred) to be trained in statistic during your...''}. Respondents could check any of the options in the legend, for each stage of their career.}
    \label{fig:q10}
\end{figure}

\textbf{Figure~\ref{fig:q10} displays how people would prefer to be trained in statistics at different stages in their educational and academic career, based on the responses to Question 10.} At the undergraduate stage, people would prefer taking a course in either the astro/physics department or the math/stats department. However, once people reach the graduate level, they would prefer to learn about statistical methods within their own department. At the postdoctoral level, the desire to take courses in statistical methods drops dramatically, but the preference for workshops stays roughly the same. The relative preference for mentoring and meetings with statisticians at the postdoc level compared to courses is also much higher than at the undergrad and graduate level.

\begin{figure}
    \centering
    \includegraphics[scale=0.56, trim={0cm 1 0 3}, clip]{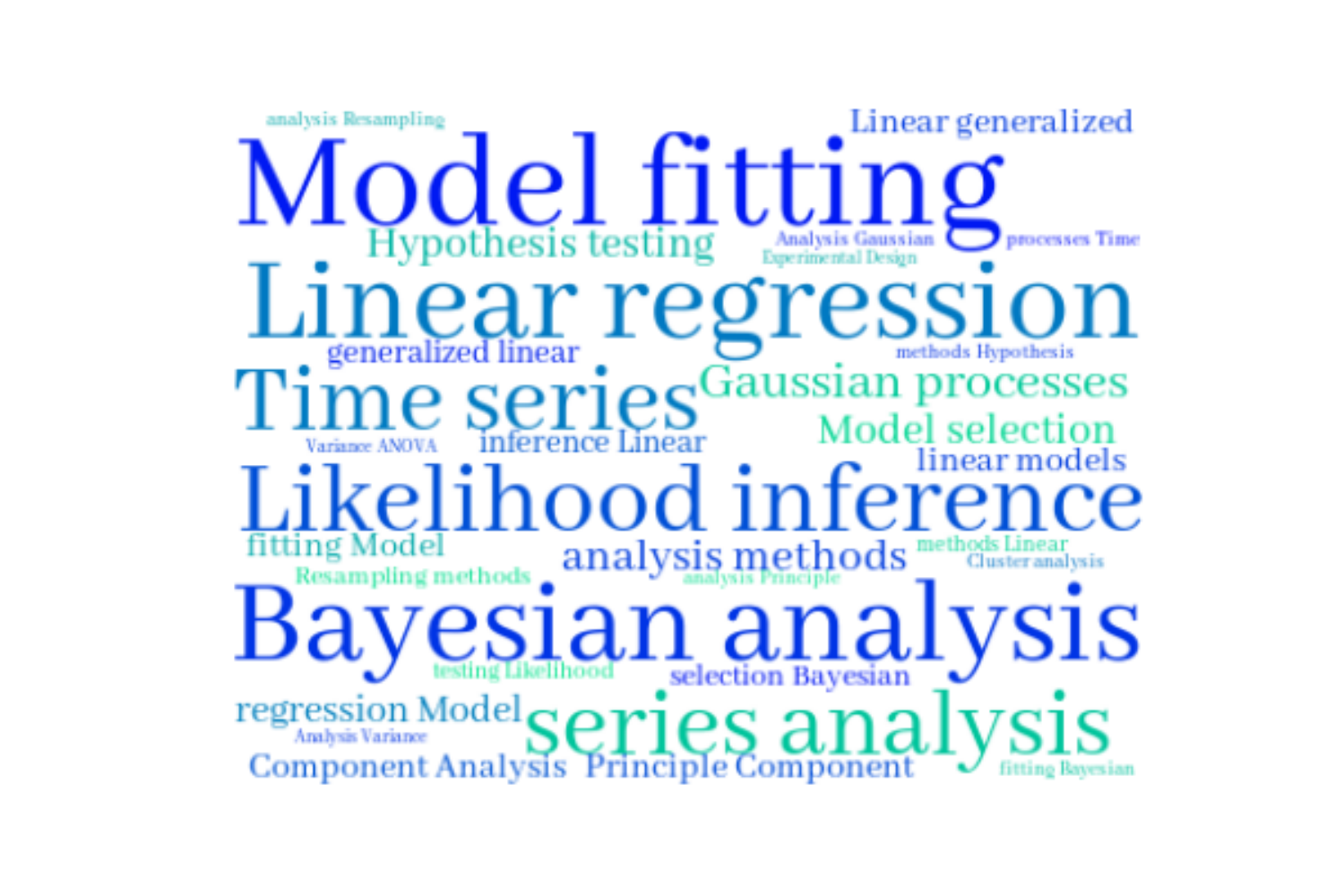}
    \includegraphics[scale=0.56, trim={0cm 1 0 3}, clip]{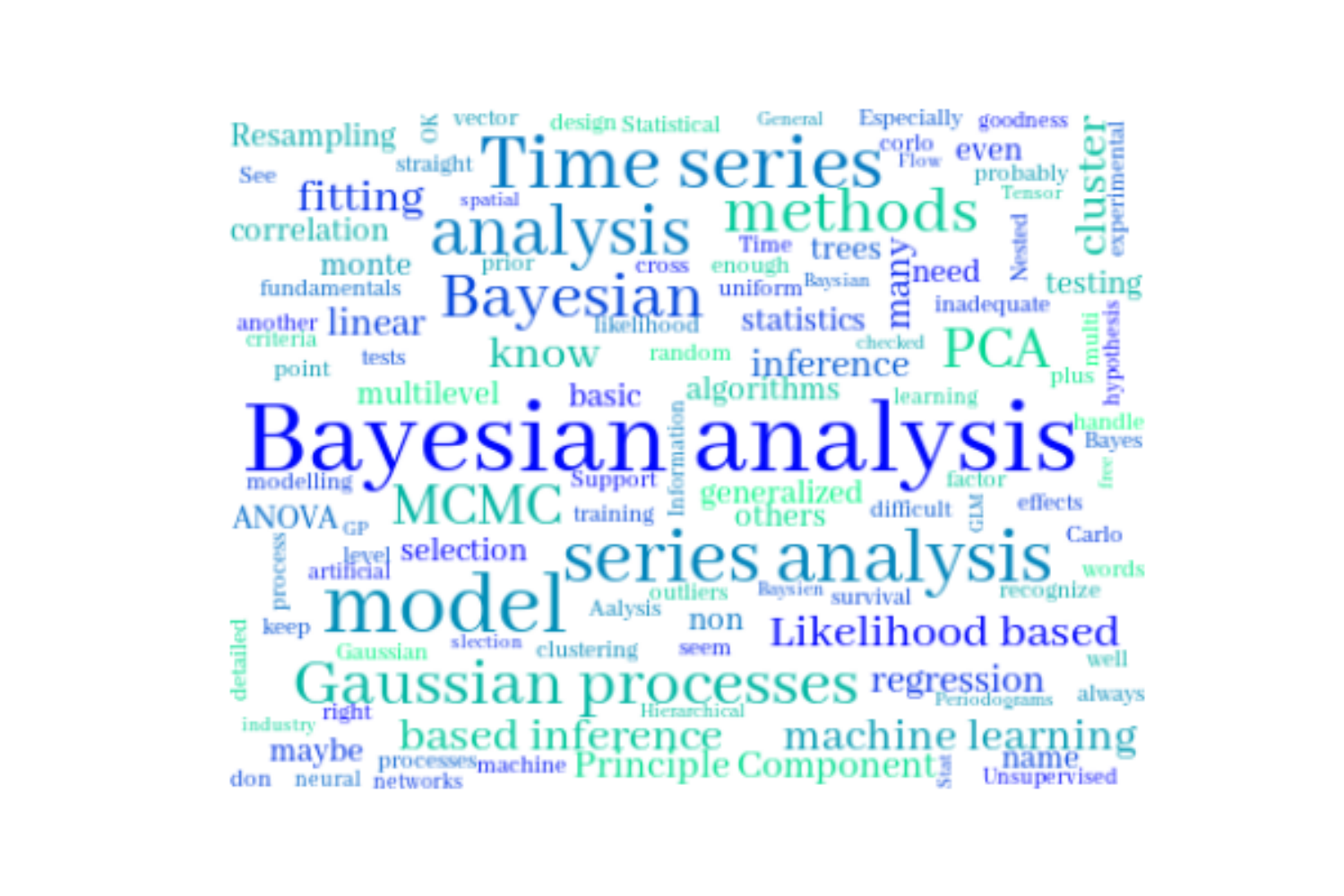}
    \vspace{-3ex}
    \caption{Left: Wordcloud generated from responses to the question \emph{``What are the statistical tools/models you find useful in your work?''}. Right: Wordcloud generated from responses to the question \emph{``What statistical tools/models do you wish you knew more about?''}}
    \label{fig:wordclouds}
    \vspace{-1ex}
\end{figure}

Questions 11 and 12 focused on the types of statistical tools and methods astronomers find useful for their work and what tools and methods they wish they knew more about, respectively. For Question 11, respondents could choose from a list of choices, but there was also a write-in ``other'' category. Question 12 was completely open-ended, but respondents could use any of the terms mentioned in Question 11 as well as add their own. Wordclouds\footnote{generated using the Python \texttt{WordCloud} package:  \url{https://amueller.github.io/word_cloud/}} are a nice way to visually represent the answers to these questions because the size of the words are proportional to the number of occurrences in the data (Figure~\ref{fig:wordclouds}). Methods such as model fitting, linear regression, likelihood inference, and Bayesian analysis appear to be some of the most important methods for astronomers. At the same time, Bayesian analysis is a topic that many astronomers wish they knew more about.

\begin{wrapfigure}{R}{0.45\textwidth}
\vspace{-5ex}
    \begin{center}
        \includegraphics[scale=0.56, trim={1cm 2 1 1}, clip]{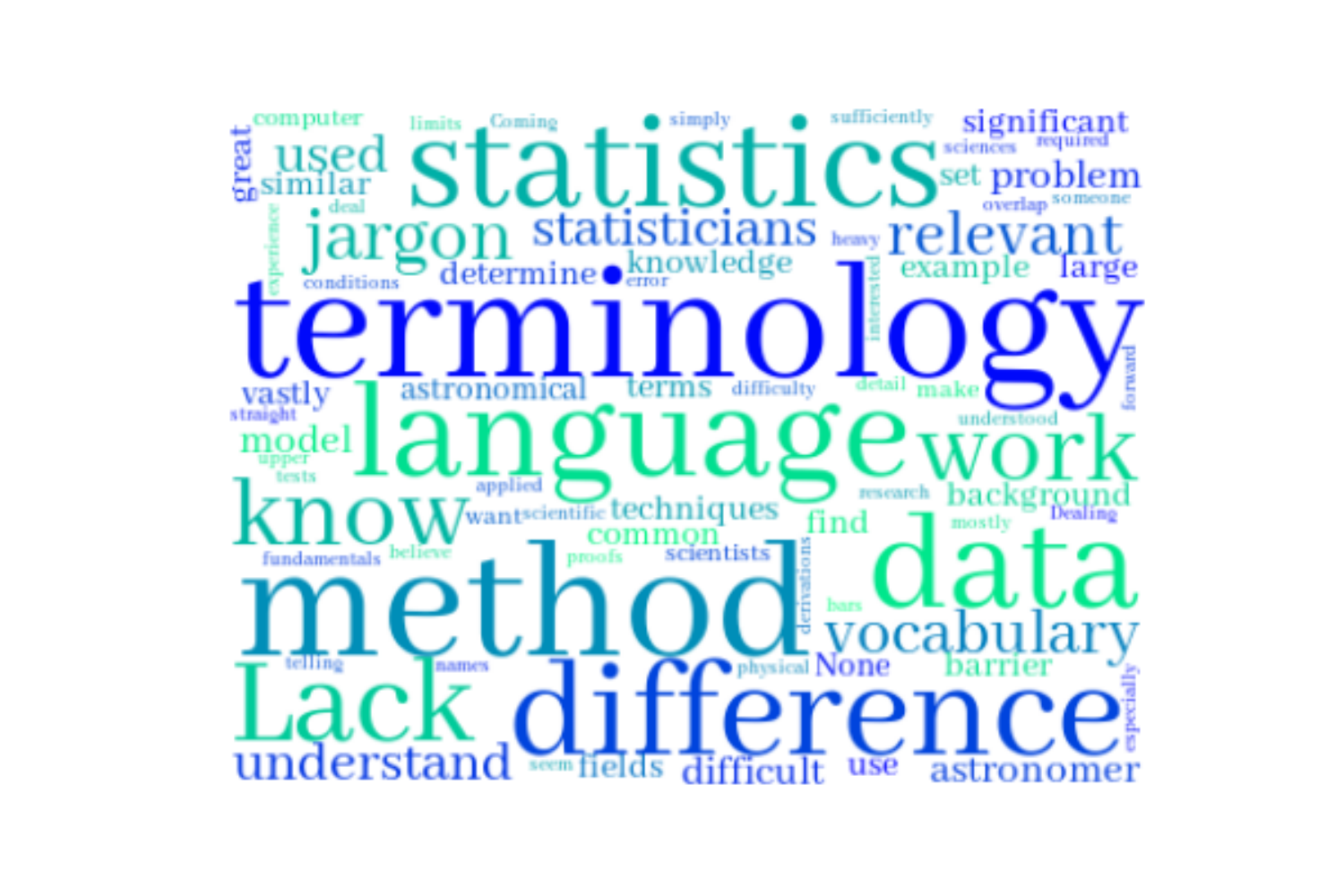}
    \end{center}
    \vspace{-3ex}
        \caption{Wordcloud created from the responses to Question 14: \emph{``What are the main challenges you have encountered when working with statisticians, applied mathematicians, or computer scientists?''}}
        \label{fig:challenges}
        \vspace{-3ex}
\end{wrapfigure}

Questions 13, 14, and 15 focused on interdisciplinary collaboration. The responses to Question 13 revealed that approximately 63\% of respondents have never collaborated with a statistician, computer scientist, or applied mathematician. A summary of the responses to Question 14 is displayed through a wordcloud (Figure~\ref{fig:challenges}). Respondents were asked to list the main challenges when working with statisticians, applied mathematicians, or computer scientists. The words that repeatedly showed up were ``terminology'', ``method'', ``language'', and ``difference''. Question 15 asked what the biggest barriers were to interdisciplinary collaboration, and overwhelmingly people mentioned the word ``time''.

The final question of the survey (Question 16) asked for open comments in regards to statistics training of astronomers and/or astrostatistics in Canada. As these are difficult to quantify, we instead include some examples in the next section to augment the discussion. 



\section{Discussion}\label{sec:discussion}

\textbf{The survey responses reflect the striking gap between what astronomers and astrophysicists want and need in terms of statistical training and what they are actually provided during their formative years.} Many respondents expressed that statistical analyses are very important or even critical to their research, but that they also (1) lacked formal training in statistics, (2) had difficulty with statistics terminology, and (3) wished for more formal training during undergraduate and graduate school.

Many of the free-form answers to the final question of the survey also reflect these findings. Respondents clearly stated their wishes for formal training in statistics --- here are just a few examples:

\begin{quote}
    \emph{- I would be a much stronger researcher, and be able to produce much more meaningful and exciting science, if I had received proper training in statistics. Instead, I had a series of upper undergrad and graduate courses that *assumed* we knew the basics of stats, but never actually taught us any of it. I've only had one formal data analysis course, and because our education had been so lacking up to that point, the course tried to fit too much into too little time. I still would not be confident doing any serious statistical analysis.}

    \emph{- When I was in grad school, myself and all my classmates wish we we had taken some sort of course in data analysis or statistics before graduate school. Many of us learned only limited data analysis and statistics in a physics lab or astronomy course in undergrad, and we did not feel equipped for the data analysis challenges associated with actual research.}

    \emph{- I wish I had training in statistics much earlier in my career.}
    
    \emph{- I really wish I had at least one class of statistic and it would be wonderful to have a astrostatistic class!}

    \emph{- Undergraduate training in statistics is severely lacking in most astro/physics departments I have heard about, and fixing that would do so much to help new astronomers in the field.}
\end{quote}

As this particular respondent points out, statistical analysis is important for their work in industry, too:
\begin{quote}
    \emph{- I had zero exposure to stats in physics and astrophysics. I took one undergraduate course in an engineering department, but most of my stats skills are learned in industry when needed. This has been my biggest challenge in industry. Stats are extremely important in analyses and making informed business decisions and should be emphasized much, much more in physics departments!}
\end{quote}
As one-half to two-thirds of astronomy Ph.D. recipients do not find employment in academic astronomy jobs \citep{Metcalfe2008}, gaining statistical skills related to data analysis is important for students to be highly employable in data science industry jobs. Thus, more focused training in statistics of students will improve our training of HQP in Canada --- it will benefit both students who want to stay in academia and those who want to pursue quantitative (e.g. data scientist) careers elsewhere. 


There was also an element of hope in many comments, suggesting the Canadian astronomical community would be amiable to many of the recommendations we give in Section~\ref{sec:recs}. Below are a few examples of such comments:

\begin{quote}
    \emph{- Definitely something that is going to be required A LOT in the very near future.}

    \emph{- Training (ie: courses and/or workshops) for astronomers would be great to be more available. It doesn't necessarily need to be hard, but practice and getting the fundamentals understood clearly and solidly would be excellent.}
    
    \emph{- I think it's great that this is becoming a recognized important skillset in astronomy and hope that we are able to improve our level of training for upcoming students \& postdocs}
    
    \emph{- My university offers a graduate course in Statistics in Astronomy and I encourage all my grad students to take it}
    
\end{quote}


In general, the survey results suggest many Canadian astronomers desire improved training in statistics and data analysis methods. Below, we list actionable recommendations that can be taken at the institutional, national, and international level. If acted upon, these recommendations should improve the statistical training of HQP in astronomy, increase interdisciplinary collaboration in astronomy research, provide future opportunities for astrostatistics research in Canada, and help make Canada a leader in the field of astrostatistics.

\section{Recommendations \& Proposed Initiatives}\label{sec:recs}

\subsection{Institutional Level}

\begin{itemize}[itemsep=2pt]
    \item Add or change at least one course to the undergraduate astronomy and physics program to incorporate training in applied statistical and exploratory data analysis. Ideally, this should be done in collaboration with statistics departments.
    
    \item Create venues where graduate students and postdocs in departments of physics \& astronomy can meet researchers in methods-focused departments such as statistics, computer science, and applied mathematics, and vice versa.
    
    \item Initiate joint hires between department of statistics and astronomy/physics. Such joint hires will catalyze collaborative efforts, will raise awareness in both fields, and will transform the graduate education experience for multiple students at a time.
    
    \item Create research groups within your institution that span both disciplines. While standard analyses can be conducted with basic statistical knowledge, past experiences show that complex problems arise in terms of model formulation and/or computation, and in terms of data integration techniques. Both of these require close collaboration between statisticians and astronomers/astrophysicists.
    %
    
    \item In addition to joint hires, reward interdisciplinary work done by faculty, for example through faculty progression and nomination for awards.
    
\end{itemize}

\subsection{National Level}

\begin{itemize}
    
    \item Create organizational chapters and/or working groups (WGs) in Astrostatistics in the Statistical Society of Canada (SSC) and in the Canadian Astronomical Society (CASCA). These chapters/WGs could ensure that each annual meeting has scientific sessions dedicated to research in astrostatistics and astroinformatics. They could also organize workshops and conferences in collaboration with organizations in the US and Europe. 
    
    \item \textbf{Collaborate with the Canadian Statistical Sciences Institute (CANSSI)} 
    \begin{itemize}
    
        \item CANSSI has committed to developing curricula, and when necessary, online courses in emerging areas of statistical and data science. \textbf{The astrostatistics community should work with CANSSI to develop a curriculum specifically geared towards astronomers.}  The current CANSSI Director has indicated a willingness to host a workshop at SFU to help lead this endeavour.
        
        \item Propose interdisciplinary workshops and conferences in astrostatistics through the call for Emerging Fields in Statistical Sciences
        
        \item \textbf{Develop CANSSI Collaborative Research Team Projects, which provide funding for three years to statistical scientists and collaborators from other disciplines\footnote{\url{http://www.canssi.ca/canssi-collaborative-research-team-projects/})}.} This venue could provide an excellent opportunity for building collaborations between astronomers and statisticians across the country.
        
        \item Astronomers should also partner and work with CANSSI at the provincial level, as there are regional CANSSI nodes across the country. For example, the mandate of CANSSI-Ontario at the University of Toronto is to stimulate statistical sciences-related work in the province. In this context, CANSSI-Ontario can facilitate and encourage establishing ties between data scientists and astronomers in Ontario, and also nationally, to accelerate the development of astrostatistics.
        
    \end{itemize} 
        
    \item Raise interest for astronomy-related problems among statisticians, for example through a special issue on Astrostatistics for the Canadian Journal of Statistics (CJS)
    
    \item Create a set of guidelines for interdisciplinary conferences and workshops. For example, (1) make sure to include a mix of statisticians and astronomers on the LOC and in the program, (2) clearly identify intended learning outcomes for attendees while organizing workshop programs, and (3) stress to presenters the importance of defining all terminology clearly, and not assuming ``basic'' knowledge in astronomy or statistics.
    \item Put forward regularly proposals for astrostatistics workshops or thematic programs at Banff's BIRS, CRM in Montreal, and Fields Institute in Toronto.
    
    \item Enhance the Canadian Advanced Network For Astronomy Research (CANFAR) cyber-platform to deliver packages and environments for statistical analysis in astronomy using interfaces and approaches familiar to that research community.  Such a platform could provide tutorials or example workflows relevant to the data sets stored within CANFAR.
    
    \item As a long-term goal, set up a national-level Astrostatistics \& Astroinformatics Centre of Excellence. A centre that spanned institutions across Canada would act as a natural catalyst to engaging astronomers and statisticians leading to a more common understanding of the terminology and problem spaces within this multi-disciplinary realm. Funding programs like NSERC-CREATE could initiate the establishment of such a centre, and such a program would be a natural partner for Canada's Networks of Centres of Excellence. 
    
    
    
\end{itemize}


\vspace{1ex}
\subsection{International Level}

\begin{itemize}
\itemsep0em
    \item With the formation of a CASCA astrostatistics chapter or WG, executive members should collaborate with other organizations (AAS WGAA, ASA AIG, IAU B3 Commission, etc.) in ways that will benefit both Canadian-based astronomers and the international community. For example, a conference on astrostatistics and astroinformatics that is open to international participants could be hosted in a major Canadian city and/or University.
    
    \item Advocate for and help develop statistical standards for astronomy, similar to other fields, especially in regards to developing publication policies for statistical requirements in major journals.

    \item Help develop a standard curriculum in astrostatistics at the graduate level through collaboration with international organizations.

    \item Capitalize on the momentum of astrostatistics and astroinformatics and leverage the international connections we already have to recruit postdocs and potential faculty who are  trained in these areas.
    
\end{itemize}

\vspace{2ex}
\section{Closing Remarks}

 \textbf{Given the plethora of astrostatistics and astroinformatics associations and activites outside of Canada, Canadian astronomy has an opportunity and obligation to strengthen our training of HQP and our research output in the area of astrostatistics and astroinformatics}. We believe the existing gap between training and post-graduate professional development can be closed by building strong ties between astro and stats communities at institutional and national levels. Similar ties have been already developed in the UK and US resulting in the first wave of PhD graduates that are genuinely interested in research cross-fertilisation. Canadian organizations have the opportunity to recruit some of these astrostatisticians and to encourage the formation  of  interdisciplinary teams around them.
 


\newpage
\section{LRP Criteria Questions}

\begin{lrptextbox}[How does the proposed initiative result in fundamental or transformational advances in our understanding of the Universe?]

Increasing collaboration between statisticians and astronomers will improve our inference capabilities of difficult problems in astrophysics and cosmology. In particular, our recommendations will enable us to make the most of the data we have. Astrostatistics and astroinformatics research also informs and improves simulation studies, which allow us to better understand the Universe where we do not have enough observations.




\end{lrptextbox}

\begin{lrptextbox}[What are the main scientific risks and how will they be mitigated?]

Increasing the statistical training of astronomers in Canada, and fostering more research collaborations between statisticians and astronomers has limited scientific risk. \textbf{The biggest scientific risk would be to ignore the global push towards more statistical training and leadership --- if we do not act now, we run the risk of being left behind internationally.} Acting on the recommendations in this white paper will make us more competitive internationally. Astrostatistics requires open communication between the statistics and astronomy communities. Although interdisciplinary collaboration takes time to develop and changing curriculum requirements requires working through bureaucratic processes (both of which might be seen as a ``risks''), the benefits and long-term payoff in terms of research quality and training of HQP, in our opinion, outweigh these risks. 

\end{lrptextbox}

\begin{lrptextbox}[Is there the expectation of and capacity for Canadian scientific, technical or strategic leadership?] 

Canadian universities hired two astrostatisticians in 2019; we have the unique opportunity to grow a program in astrostatistics and become a leading country in this field. Moreover, Canada has some of the world's leading statisticians. In 2018, the new national headquarters of the CANSSI at Simon Fraser University's Big Data Hub opened, as well as regional nodes such as CANSSI-Ontario in Toronto. One of CANSSI's goals is to foster interdisciplinary collaboration between statisticians and domain sciences --- the astronomical community should jump on this opportunity.

\end{lrptextbox}

\begin{lrptextbox}[Is there support from, involvement from, and coordination within the relevant Canadian community and more broadly?] 

Yes, statisticians in Canada are starting to become aware of the field of astrostatistics (some are co-authors on this white paper) and CANSSI will likely be excited to help initiate collaborations with astronomers. More broadly, there is international support for our recommendations from the WGAA and the ASA AIG in the US.

\end{lrptextbox}

\begin{lrptextbox}[Will this program position Canadian astronomy for future opportunities and returns in 2020-2030 or beyond 2030?] 

Yes, improving training and increasing astrostatistics and astroinformatics research in Canada will equip astronomers with skills that are in demand for many national and international astronomy research programs, both in theory and observation. In particular, HQP with skills in statistical analysis and computing are and will be highly sought after for projects with Big Data, such as ALMA, LSST, CHIME, Euclid, and SKA.  For example, Canada will be leading the world in fast radio burst data with the CHIME telescope and we want to make the best use of these data as possible.

\end{lrptextbox}

\begin{lrptextbox}[In what ways is the cost-benefit ratio, including existing investments and future operating costs, favourable?] 

The cost-benefit ratio is extremely favourable; by investing in improved training in statistical methods and collaborating with statisticians, Canadian-trained astronomers and Canadian-based interdisciplinary collaborations will make the most efficient use of data generated and collected through Canadian-funded projects. Because statistical analysis is essential in both academia and industry, astrostatistics research may seek funding through industry partnerships (e.g., Mitacs).

\end{lrptextbox}

\begin{lrptextbox}[What are the main programmatic risks and how will they be mitigated?]

Programmatic risks exist at the institutional, national, and international level (although the latter are less easily mitigated by Canadian-based astronomers and statisticians alone). At the institutional level, intellectual buy-in is necessary for interdisciplinary training and research to be successful. At the provincial and national levels, barriers may include obtaining funding from agencies such as OGS and NSERC, which traditionally fund projects defined within certain evaluation groups. However, this is changing and there are avenues of NSERC funding that specifically promote advances and innovation through interdisciplinary work.  

\end{lrptextbox}

\begin{lrptextbox}[Does the proposed initiative offer specific tangible benefits to Canadians, including but not limited to interdisciplinary research, industry opportunities, HQP training, EDI, outreach or education?]


 Yes. The proposed recommendations and initiatives offer tangible benefits to interdisciplinary research, industry opportunities, HQP training, and education. HQP with strong statistical analysis and computing skills will be highly desirable in both academia and industry. Statistical methodologies developed through research in astrostatistics can have broader impacts in other disciplines and industries that are also trying to answer scientific questions with big data.


\end{lrptextbox}

\bibliography{example} 

\begin{thebibliography}{}
\expandafter\ifx\csname natexlab\endcsname\relax\def\natexlab#1{#1}\fi

\bibitem[{Babu \& Feigelson(2012)}]{babu2012statistical}
Babu, G.~J., \& Feigelson, E.~D. 2012, Statistical Challenges in Modern
  Astronomy II (Springer Science \& Business Media)

\bibitem[{{Borne} {et~al.}(2009){Borne}, {Accomazzi}, {Bloom}, {Brunner},
  {Burke}, {Butler}, {Chernoff}, {Connolly}, {Connolly}, \&
  {Connors}}]{borne2009}
{Borne}, K., {Accomazzi}, A., {Bloom}, J., {et~al.} 2009, in astro2010: The
  Astronomy and Astrophysics Decadal Survey, Vol. 2010, P6

\bibitem[{{Eadie} {et~al.}(2019){Eadie}, {Loredo}, {Mahabal}, {Siemiginowska},
  {Feigelson}, {Ford}, {Djorgovski}, {Graham}, {Ivezic}, {Borne},
  {Cisewski-Kehe}, {Peek}, {Schafer}, {Yanamandra-Fisher}, \&
  {Young}}]{2019arXivEadie}
{Eadie}, G., {Loredo}, T.~J., {Mahabal}, A.~A., {et~al.} 2019, arXiv e-prints,
  arXiv:1909.11714

\bibitem[{Feigelson \& Babu(2012)}]{feigelson2012modern}
Feigelson, E.~D., \& Babu, G.~J. 2012, Modern statistical methods for
  astronomy: with R applications (Cambridge University Press)

\bibitem[{Ford \& Gregory(2007)}]{ford2007statistical}
Ford, E., \& Gregory, P. 2007, in ASP Conf. Ser, Vol. 371, 189

\bibitem[{Hilbe(2012)}]{hilbe2012astrostatistical}
Hilbe, J.~M. 2012, Astrostatistical Challenges for the New Astronomy, Vol.~1
  (Springer Science \& Business Media)

\bibitem[{Hilbe {et~al.}(2017)Hilbe, De~Souza, \& Ishida}]{hilbe2017bayesian}
Hilbe, J.~M., De~Souza, R.~S., \& Ishida, E.~E. 2017, Bayesian models for
  astrophysical data: using R, JAGS, Python, and Stan (Cambridge University
  Press)

\bibitem[{{Hlo{\v{z}}ek}(2019)}]{Hlozek2019PASP}
{Hlo{\v{z}}ek}, R. 2019, \pasp, 131, 118001

\bibitem[{Ivezi{\'c} {et~al.}(2014)Ivezi{\'c}, Connolly, VanderPlas, \&
  Gray}]{ivezic2014statistics}
Ivezi{\'c}, {\v{Z}}., Connolly, A.~J., VanderPlas, J.~T., \& Gray, A. 2014,
  Statistics, data mining, and machine learning in astronomy: a practical
  Python guide for the analysis of survey data, Vol.~1 (Princeton University
  Press)

\bibitem[{{Loredo} {et~al.}(2009){Loredo}, {Accomazzi}, {Bloom}, {Borne},
  {Brunner}, {Burke}, {Butler}, {Chernoff}, {Connolly}, \&
  {Connolly}}]{loredo2009}
{Loredo}, T.~J., {Accomazzi}, A., {Bloom}, J., {et~al.} 2009, in astro2010: The
  Astronomy and Astrophysics Decadal Survey, Vol. 2010, P34

\bibitem[{{Metcalfe}(2008)}]{Metcalfe2008}
{Metcalfe}, T.~S. 2008, \pasp, 120, 229

\bibitem[{{Siemiginowska} {et~al.}(2019){Siemiginowska}, {Eadie}, {Czekala},
  {Feigelson}, {Ford}, {Kashyap}, {Kuhn}, {Loredo}, {Ntampaka}, {Stevens},
  {Avelino}, {Borne}, {Budavari}, {Burkhart}, {Cisewski-Kehe}, {Civano},
  {Chilingarian}, {van Dyk}, {Fabbiano}, {Finkbeiner}, {Foreman-Mackey},
  {Freeman}, {Fruscione}, {Goodman}, {Graham}, {Guenther}, {Hakkila},
  {Hernquist}, {Huppenkothen}, {James}, {Law}, {Lazio}, {Lee},
  {L{\'o}pez-Morales}, {Mahabal}, {Mandel}, {Meng}, {Moustakas}, {Muna},
  {Peek}, {Richards}, {Portillo}, {Scargle}, {de Souza}, {Speagle}, {Stassun},
  {Stenning}, {Taylor}, {Tremblay}, {Trimble}, {Yanamand ra-Fisher}, \&
  {Young}}]{Siemiginowska2019}
{Siemiginowska}, A., {Eadie}, G., {Czekala}, I., {et~al.} 2019, \baas, 51, 355

\end{thebibliography}


\end{document}